\documentclass[conference]{IEEEtran}
\IEEEoverridecommandlockouts
\usepackage{cite}
\usepackage{amsmath,amssymb,amsfonts}
\usepackage{algorithmic}
\usepackage{graphicx}
\usepackage{textcomp}
\usepackage{xcolor}

\usepackage{multirow}
\usepackage{listings}
\usepackage{hyperref}
\usepackage{color}
\usepackage{threeparttable}

\hypersetup{
    colorlinks=true,
    linkcolor=blue,
    citecolor=blue,
    urlcolor=blue,
}

\def\BibTeX{{\rm B\kern-.05em{\sc i\kern-.025em b}\kern-.08em
     T\kern-.1667em\lower.7ex\hbox{E}\kern-.125emX}}
\begin{document}

\title{PMU-Data: Data Traces Could be Distinguished}
% \author{Anonymous Submission}
\author{
\IEEEauthorblockN{
    Zhouyang Li\IEEEauthorrefmark{2}\IEEEauthorrefmark{3}, 
    Pengfei Qiu\IEEEauthorrefmark{2}\IEEEauthorrefmark{3}$^*$\thanks{$^*$Corresponding author},
    Yu Qing\IEEEauthorrefmark{3}
    Chunlu Wang\IEEEauthorrefmark{2},
    Dongsheng Wang\IEEEauthorrefmark{3}\IEEEauthorrefmark{4},
    Xiao Zhang\IEEEauthorrefmark{5},
    Gang Qu\IEEEauthorrefmark{6}
    }
    \\
\IEEEauthorblockA{
    \IEEEauthorrefmark{2}Key Laboratory of Trustworthy Distributed Computing and Service (BUPT), Ministry of Education, Beijing, China\\
    \IEEEauthorrefmark{3}Zhongguancun Laboratory, Beijing, China~~
    \IEEEauthorrefmark{4}Tsinghua University, Beijing, China\\
    \IEEEauthorrefmark{5}Beijing University of Technology	~~
    \IEEEauthorrefmark{6}University of Maryland, College Park, MD, USA
    }
\IEEEauthorblockA{
    \{li\_zhouyang1,qpf\}@bupt.edu.cn, qingyu@zgclab.edu.cn, wangcl@bupt.edu.cn \\ wds@tsinghua.edu.cn, zhangxiao@bjut.edu.cn, gangqu@umd.edu
    }
}

\maketitle

\begin{abstract}
Modern processors widely equip the Performance Monitoring Unit (PMU) to collect various architecture and microarchitecture events. Software developers often utilize the PMU to enhance program's performance, but the potential side effects that arise from its activation are often disregarded. In this paper, we find that the PMU can be employed to retrieve instruction operands. Based on this discovery, we introduce PMU-Data, a novel category of side-channel attacks aimed at leaking secret by identifying instruction operands with PMU.

To achieve the PMU-Data attack, we develop five gadgets to encode the confidential data into distinct data-related traces while maintaining the control-flow unchanged. We then measure all documented PMU events on three physical machines with different processors while those gadgets are performing. We successfully identify two types of vulnerable gadgets caused by \texttt{DIV} and \texttt{MOV} instructions. Additionally, we discover 40 vulnerable PMU events that can be used to carry out the PMU-Data attack. We through real experiments to demonstrate the perniciousness of the PMU-Data attack by implementing three attack goals: (1) leaking the kernel data illegally combined with the transient execution vulnerabilities including Meltdown, Spectre, and Zombieload; (2) building a covert-channel to secretly transfer data; (3) extracting the secret data protected by the Trusted Execution Environment (TEE) combined with the Zombieload vulnerability.
\end{abstract}


%%%% 7. PAPER CONTENT %%%%
\section{INTRODUCTION}
\label{sec:introduction}
Modern processors are seriously threatened by a set of side-channel attacks, which are mostly caused by the contention of the shared resources such as caches~\cite{percival2005cache,liu2015last}, scheduler queue~\cite{gast2022squip}, retirement~\cite{xu2024retirement}, et al. There are generally three steps to achieve side-channel attacks: 1) the attacker prepares the shared resource; 2) the victim leaves secret-related content on the resource; 3) the attacker speculates the secret from the resource. In this paper, we propose a new type of side-channel attack that does not rely on any contention of shared resources. It is caused by the inherent feature of instructions and the Performance Monitoring Unit (PMU)~\cite{Intel2023sdm}.
% reorder buffer~\cite{ap2021RoBcontention},

The PMU is a significant processor module, which provides several counters to track instruction execution by monitoring events (called PMU events) like instruction cycles, memory loads, retired instructions. Besides, existing studies~\cite{qiu2023spill} have verified that PMU is capable of monitoring the events triggered during transient executions (the instructions are performed but the execution results are not submitted because of some special circumstances such as an exception that is induced in out-of-order execution~\cite{lipp2018meltdown}, the prediction that is incorrect in speculation execution~\cite{kocher2019spectre}, or a microcode-assist execution that occurs in Microarchitectural Data Sampling (MDS)~\cite{schwarz2019zombieload}). Above all, PMU is shared during both transient and non-transient executions, and its value is permanent.
%In this paper, we perceive the transient window as the victim field.  PMU is a significant hardware module of current processors, which provides several registers (called PMU counters) to depict the execution process of instructions by monitoring specific events (PMU events) in processors, such as instruction cycles, memory loads, retired instructions, etc.

Data traces on the PMU primarily reflect the function of the instruction rather than its operands. For instance, with the exclusive-or (\texttt{XOR}) instruction, the PMU captures consistent events since the operands do not introduce any variability in the trace. However, certain instructions exhibit unique behaviors when encountering specific operands. For instance, a division (\texttt{DIV}) instruction that encounters a dividend of \texttt{zero} will trigger an exception, halting the execution of the instruction, which can be captured by the PMU event \texttt{ARITH.DIVIDER\_ACTIVE}. This event means how many cycles when the divide unit is busy executing divide or square root operations. Then, we can identify whether the dividend of a \texttt{DIV} instruction is \texttt{zero} with this PMU event.
%This gives us a chance to construct a side-channel attack by tracing the operand of the instructions in transient executions based on the data traces left within the PMU counters.

Based on this discovery, we propose the PMU-Data attack, which is a new side-channel attack that leaks the secret data in transient executions from the data traces on PMU. We manually analyze instructions, and propose a novel mechanism to locate PMU side-channel attacks. Different solutions to those challenges are described in this paper. 
%(移到后面了)Moreover, combined with two facts, the PMU-Data attack is powerful. (1) When implementing Meltdown and Zombieload, the PMU-Data attack can be realized using the signal mechanism (\verb|setjmp()| function) to handle exceptions. (2) Unprivileged attackers could access PMU values using transient execution vulnerabilities \cite{Weber2023Reviving}. Therefore, our PMU-Data attack does not rely on the support of Transactional Synchronization Extensions (TSX) and root privilege, but threat many processors and scenarios.

\begin{figure*}[htbp]
\begin{center}
\includegraphics[width=1.9\columnwidth]{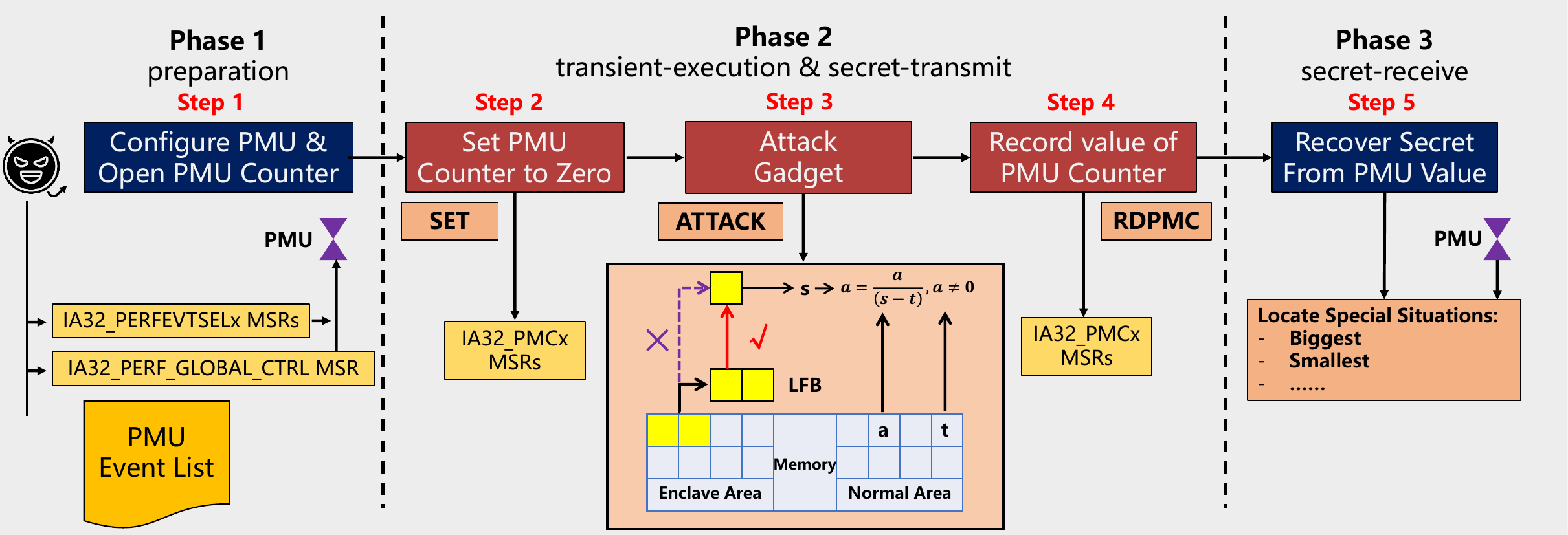}
\caption{An overview of PMU-Data attack}
\label{fig:overview}
\end{center}
\end{figure*}

In i7-6700, i7-7700, and i5-7300U Intel processors, we demonstrate that the PMU-Data attack can be utilized to implement Meltdown, Spectre, and ZombieLoad attacks. Besides, we evaluate the function of PMU-Data to serve as a covert channel. Specifically, in i7-7700, we successfully leak the SGX-protected secret data. The Intel SGX provides hardware support for Trusted Execution Environment and protects code and data from modification by privileged attackers. When we use PMU-Data to implement Zombieload in order to steal the SGX-protected secret, experiment results show that the throughput of the PMU-Data attack can be up to 76 KB per second with an average error rate of 0.33\%.
%for PMU-Data attack.

\noindent \textbf{Contributions.}
%Based on our discovery that PMU could disclose the data traces left by some special instructions while processing different operands, we make the following contributions:
\begin{itemize} 
%（写三个，合并）这个40显得很奇怪
    \item We discover that PMU can disclose the data traces left by some special instructions while processing different operands. Based on it, we propose PMU-Data attack, a new side-channel attack that leaks secret data by recovering the operands of instructions with PMU.
    \item We classify the Intel x86 instructions into five categories to find out two kinds of vulnerable gadgets, and evaluate the PMU events on i7-6700, i7-7700, i5-7300U Intel processors to locate 40 vulnerable PMU events.
    \item We successfully use PMU-Data to implement Meltdown, Spectre and Zombieload to leak the secret (including the SGX-protected secret data). And we successfully use PMU-Data to build a covert-channel.
\end{itemize}
    % \item We propose the PMU-Data attack, which is a new side-channel attack that leaks the secret data by recovering the operands of instructions with PMU.
    % \item We analyze the Instruction Set Architecture (ISA) of Intel X86, and classify those instructions into five categories. In experiment, we construct several instruction gadgets to encode the secret data into PMU events, in which the secret data and normal data are fed to the same instruction but they will leave different data traces. The data trace that is recognized using PMU exposes whether the operand of the instruction is secret data or normal data.

%\noindent \textbf{Outline.} In \autoref{sec:background}, we introduce PMU side-channel attacks and transient execution attacks. In \autoref{sec:overview}, we provide a high-level overview of the proposed PMU-Data side-channel attack. In \autoref{sec:case}, we provide two variants realized in this work, and analyze the strengths of our PMU-Data attack. In \autoref{sec:experiment}, we display the experiment results on physical machines. Moreover, we provide the effective mitigation in \autoref{sec:mitigation+discuss}.

\section{BACKGROUND}
\label{sec:background}
\subsection{Performance Monitoring Unit (PMU)}

Performance Monitoring Unit~\cite{Intel2023sdm} is actively used to optimize applications by measuring parameters such as cache misses, machine clears, branch misses, etc. However, the PMU can also be utilized to breach the secure boundary of processors. For example, PMU-Spill
~\cite{qiu2023spill} discovers that special situations of the PMU values take place when results of the \texttt{if} instruction are different (taken or not-taken).

\subsection{Transient Execution Attack}
To utilize different execution units in parallel, an instruction stream is decoded into multiple micro-operations, which share execution units dynamically within sibling threads. The microarchitectural state was long considered non-observable. However, Meltdown~\cite{lipp2018meltdown}, Spectre~\cite{kocher2019spectre}, and other transient execution attacks~\cite{schwarz2019zombieload} have demonstrated that hardware should not be completely trusted.
\section{OVERVIEW OF PMU-Data ATTACK}
\label{sec:overview}
% In this section, we present our motivation to propose the PMU-Data side-channel attack, and design a threat model.
\subsection{Motivation}
Performance Monitoring Unit (PMU) can be an excellent side-channel, due to its inherent feature of shared between transient and non-transient executions. Therefore, we do experiments to research whether PMU could reveal the operands of instructions in transient executions. We categorize the Intel x86 instructions into five classes, and construct gadgets to locate special situations of PMU. While those gadgets are performing, because functions of different PMU events are diverse, we manually research all PMU events on three physical machines. Results show that two kinds of instructions (\texttt{DIV} and \texttt{MOV}) are vulnerable. Therefore, we propose two variants of PMU-Data attack, and evaluate them.
%We discover that special situations of PMU occur when the \verb|DIV| (division) instruction is encountering different operands. Based on this discovery, we construct a PMU side-channel attack (called PMU-Data v1). To expand our work, we categorize the instructions into five classes, and construct gadgets to search special situations of PMU. Functions of different PMU events are diverse, so we manually research all documented events in experiments. Results show that there are 39 PMU events available to build another PMU side-channel attack (called PMU-Data v2), which could recover the operands of the \verb|MOV| instruction.

\subsection{Assumption and Threat Model}
We assume the attacker aims to leak the secret data from the victim, and they are using the same Intel CPU. We do not make any special assumptions about the attacker and the victim, since CounterLeak \cite{Weber2023Reviving} shows that unprivileged attackers could also obtain the ability to read the PMU counters. In threat model, the attacker can be a root user that directly reads the value of the PMU counter, or a normal user that leaks the value of the \texttt{RDPMC} instruction by CounterLeak. The victim can be a normal application, a kernel application, or sensitive data protected by the Intel SGX. Practically, we uses the PMU-Data to recover the data in a transient window, which can bypass the cache-related and timer-related defenses, and omit the \texttt{if} instruction in PMU-Spill.
%\label{sec:method}
% In this section, we divide the process of PMU-Data into five steps, and display our method to find vulnerable instructions (\verb|DIV| and \verb|MOV|). Based on them, we design two variants, and analyze their strengths than cache side-channels and PMU-Spill.
\subsection{Attack Steps}
In \autoref{fig:overview}, we take the \texttt{DIV} instruction as the example to depict our work. In this scenario, the attacker utilizes Zombieload to read the SGX-protected secret data through the Line Fill Buffer (LFB) transiently. The PMU-Data attack recovers the secret data in a transient window. We describe more details of Step 3 in \autoref{sec:case}. With the PMU configured as \texttt{ARITH.DIVIDER\_ACTIVE} event, the counter value of special situation (when \texttt{s} equals to \texttt{t}) is smaller than other situations. From the attacker's view, there are five steps to implement the PMU-Data attack:
%We assume the size of the secret is one byte in this paper.
\begin{itemize}
    \item Step 1: It configures the \texttt{IA32\_PERFEVTSELx} MSR as a specific event, and enables the PMU function. 
    \item Step 2: It initializes the  \texttt{IA32\_PMCx} MSR.
    \item Step 3: It executes malicious gadgets to manipulate the secret and trigger special signal within the PMU value.
    \item Step 4: It reads the value of the PMU counter.
    \item Step 5: It recovers the secret data from PMU value.
\end{itemize}

\noindent \textbf{Challenges.} There are three main challenges to realize our PMU-Data attack. The first challenge (C1) is how to find vulnerable instructions that have special situations while processing different operands. The second (C2) is how to locate PMU events that can be used to identify the special situations. The third (C3) is how to recover the secret data from different values of PMU counters. We describe our solution to the first challenge here, and provide our solutions to the second and third challenges in \autoref{subsec:C2+C3}.

\subsection{Solution to C1: Instructions to Encode Secret Data into Distinct Data-Related Traces}
\label{subsec:C1} 
Some instructions leave different data-traces in PMU counters when processing different operands, which can be utilized to encode the secret data into different PMU counters in Step 3 (see in \autoref{fig:overview}). In this paper, we manually analyze the function of all instructions described in Intel$^\circledR$ 64 and IA-32 Architecture Software Developer's Manual~\cite{Intel2023sdm}. We classify these instructions into five categories, and construct gadgets for the secret data.

\textbf{Addition and Subtraction.} The finite word size limits the range of possible results of addition and subtraction operations. We take the 8 bits as the example. Here are three typical operations, $5+6$ (Trace 1, T1), $12-8$ (Trace 2, T2), and $240+100$ (Trace 3, T3). The three operations leave different traces in PMU. However, there is a special situation in T3, whose result overflows the range of \texttt{255}. Based on this, the attacker might be able to classify all traces into two categories, overflow and non-overflow. The boundary between these two categories can be regarded as a special signal.

\textbf{Division.} Dividend is divided by a divisor into an integer quotient, in which the divisor should not be zero. Here are three typical operations, $5\div2$ (T1), $5\div3$ (T2) and $5\div0$ (T3). All PMU traces left by division operations can be classified as: normal traces (such as T1, T2) and special traces (such as T3). For special traces, the divide-by-zero exception stops the execution of the divide unit. In this situation, the divide-by-zero exception can be regarded as a special signal, and captured by \texttt{ARITH.DIVIDER\_ACTIVE} PMU event.

\textbf{Multiplication.} Multiplications by constant factors are optimized as combinations of shift and addition operations. We take the operation $5\times f$ as the example, in which $f$ denotes to different factors. Here are some multiplications leaving normal traces: $5\times3$ (T1), $5\times5$ (T2), $ 5\times1$ (T3), $5\times0$ (T4), $5\times4$ (T5). The processors might complete the multiplication by renaming registers for T3. And multiplications with $f$ as the zero (T4), or a power of 2 (T5) could be optimized with just shift operations. Data trace like T3, T4 and T5 are special situations.

\begin{table}[htbp]
    \centering
    \caption{Vulnerable instructions for PMU-Data attack}
    %\scriptsize
    \footnotesize
    \begin{tabular}{c|c|c|c}
         \hline
         Instruction&\multirow{2}{*}{Particular Situation}&\multirow{2}{*}{Vulnerable} &PMU\\
         Category&&&Events\\
         \hline
         Addition&Results can overflow&No&-\\
         Subtraction&Results can overflow&No&-\\
         \textbf{Division}&\textbf{Dividend is zero}&\textbf{Yes}&\textbf{1}\\
         Multiplication&Factor is any power of 2&No&-\\
         \textbf{Data-Moving}&\textbf{Cache entries are missed}&\textbf{Yes}&\textbf{39}\\
         \hline
    \end{tabular}
    \label{tab:gadgets}
\end{table}

\textbf{Data-Moving.} Hit and miss are different situations for the cache. For example, if we prepare one area in cache, and flush just an entry, such as \verb|flush(address_A)| that is within the area. Then, we try to access \verb|address_B| that is also included in the area. In this scenario, We classify PMU traces of this operation into two categories, normal traces (if \verb|address_B != address_A|) and special traces (if \verb|address_B == address_A|). There are many PMU events to monitor this operation and do not rely on any timers. The cache-miss situation could be regarded as a special signal.

To the best of our knowledge, we are the first to consider data traces created by arithmetic operations. We devote much effort to thoroughly traverse documented PMU events, and evaluate instructions of five categories. Results show that there are two instructions (\verb|DIV| and \verb|MOV|) vulnerable to the PMU-Data attack, and 40 PMU events available to realize the PMU-Data attack. For those "safe" instructions (\verb|ADD|, \verb|SUB| and \verb|MUL|), we do not find any vulnerable events in documented PMU list, but hold a view that such instructions may be vulnerable in the future. One reason is that these instructions are good optimized choices for hardware designers. The other reason is that there are multiple hidden PMUs, and we might find vulnerable PMU events in them.

\section{CASE STUDIES}
\label{sec:case}
Based on our discovery that there are two kinds of instructions (\verb|DIV| and \verb|MOV|) vulenrable to PMU-Data attack, we propose two variants of our PMU-Data attack.
\subsection{Variant 1: Division Gadget}
\label{subsec:v1}
% 在这里可以写三个代码片段，分别说明 PMU-Data 结合 Meltdown/Zombieload/SpectreV1 的方法。

This variant is derived from the \verb|DIV| instruction, whose special situation occurs when the divisor equals to zero. We utilize the PMU-Data attack to realize three transient execution attacks (Meltdown, Spectre-PHT, Zombieload). The attacker uses the \texttt{ARITH.DIVIDER\_ACTIVE} PMU event to catch the special signal when the divisor is zero.

\begin{lstlisting}[language=c++,frame=trbl,captionpos=b, frameround = tttt,numbers = left,xleftmargin=2.5em,xrightmargin=1.2em,caption =Core gadget for PMU-Data v1, label=list:v1,morekeywords={if,maccess,xor,mov,div,movb,sub,nop,},emph={div,sub}, emphstyle=\color{red},]
// For Meltdown
    xor rax, rax
    xor rbx, rbx
    mov 0x9, rcx
    mov (rdx), rbx						
    mov (rsi), rax
    sub rbx, rax
    div rcx

// For Spectre-PHT
    if(x<array1_size)
        temp=32/(array1[x]-t);

// For Zombieload v1
    maccess(0);
    temp=32/(target[0]-t);
\end{lstlisting}

For Meltdown, \autoref{list:v1} shows that it makes subtraction between the secret data and attacker's controllable value, and feeds the computed result to the \verb|DIV| instruction as the divisor. The \verb|rdx| stores the address of attacker's controllable variable, and  \verb|rsi| stores the address of secret. If the attacker's controllable value equals to the secret, the \verb|DIV| instruction does not be executed, and the signal is monitored by PMU. 

Moreover, we display the core gadget to implement Spectre-PHT and Zombieload in \autoref{list:v1}. In Spectre-PHT, the \verb|t| denotes the attacker's controllable value, and illegal value of \verb|x| makes the \verb|array[x]| able to access the secret data. In Zombieload, the attacker flushes the area of \verb|target|, and uses illegal operation at line 15 to trigger the transient execution. The processors feed the operation at line 16 on the secret data from the LFB. The two gadgets are based on the \verb|DIV| instruction, and regard the divide-by-zero situation as the special signal in PMU event \texttt{ARITH.DIVIDER\_ACTIVE}. And the gadget for zombieload and PMU-Data v1 can be useful even in SGX-protected areas (in \autoref{table:results})

\subsection{Variant 2: Data-moving Gadget}
\label{sec:v2}

\begin{lstlisting}[language=C++,frame=trbl,captionpos=b, frameround = tttt,numbers = left,xleftmargin=2.5em,xrightmargin=1.2em,caption = Core gadgets for PMU-Data v2, label=list-v2,morekeywords={div,sub,mov,shl,movzx,maccess},emph={movzx}, emphstyle=\color{red},]
// For Meltdown
    xor rax, rax,
    mov (rsi), rax
    shl 0xc, rax
    movzx (mem, (rax), 1), rbx

// For Spectre-PHT
if(x<array1_size)
    temp=array2[array1[x]*4096];

// For Zombieload v1
    maccess(0);
    maccess(mem+target[0]*4096);
\end{lstlisting}

This variant is derived from the \verb|MOV| instruction, whose special situation occurs when the probe array misses the cache. We utilize it to achieve Meltdown, Spectre-PHT, Zombieload.

For Meltdown, \autoref{list-v2} shows that the attacker ensures the \verb|mem| area in cache, and flushes the specific entry (\verb|flush(mem+t*4096)|), in which \verb|t| denotes the attackers' controllable variable. The \verb|rsi| stores the address of the secret data. In line 5, the attacker uses the secret to request the \verb|mem| area. If the \verb|t| equals to the secret data, the \verb|MOV| instruction would trigger a cache-miss event. As for Spectre-PHT and Zombieload, vulnerable gadgets are displayed in \autoref{list-v2}.

%Then, the attackers prepare the \verb|rax| register, and attempt to access the secret data. In line 5, this instruction executes the cache operation based on the secret data. The \autoref{meltdown-v2} is similar to the core gadget of Meltdown~\cite{lipp2018meltdown}. However, the PMU-Data v2 attack does not collect left traces by probing total caches, and it just records one microarcitecture PMU event, which is triggered by the cache miss. Based on the values of PMU counters, the attackers could speculate the secret data, which is the operand of the \verb|MOV| instruction.
%\autoref{list-v2} also shows the core gadget to realize Zombieload v1 attack. This gadget merely performs a specific memory access operation. In our experiment, there are 12 PMU events able to realize the PMU-Data v2 and Zombieload when the secret data is protected by Intel SGX (in \autoref{table:zombieload-results}).
%Compared to cache side-channel attacks, the PMU-Data v2 also achieves the attack through the cache, but does not care the timing difference. It directly uses the PMU to record the cache events, and does not rely on the high-resolution timers.

\section{ATTACK IMPLEMENTATION}
\label{sec:experiment}
In this section, we do experiments on three Intel processors, and design solutions to overcome two challenges (C2 and C3). For case studies in \autoref{sec:case}, we demonstrate them on physical machines, and evaluate their functions to build a cover-channel in Simultaneous Multi-Threading (SMT) scenarios.

\subsection{Experiment Setup}
For i7-6700 and i7-7700, we successfully achieve all Meltdown, Spectre and Zombieload related cases. For i5-7300U, we implement Spectre and Zombieload using PMU-Data. Specifically, on i7-7700, we configure the Intel SGX to protect the secret data, and just replace the flush+reload gadget with PMU-Data in Zombieload's initial Proof-of-Concept. Results show that the two variants can implement Zombieload to steal the SGX-protected secret data. We just use the \verb|root| privilege to configure and read the PMU. CounterLeak has demonstrated that unprivileged attackers could leak the PMU values. Therefore, PMU-Data can be realized on user mode, and our experiments to achieve Meltdown and Spectre are useful.

\subsection{Solutions to C2+C3: Vulnerable PMU Events and How to Speculate Secret Data}
\label{subsec:C2+C3}
We describe three main challenges to implement PMU-Data attack, and provide details to overcome C1 in \autoref{subsec:C1}. For C2 and C3, because the functions of PMU events are diverse, there are no common methods. To overcome C2, we refer to the documented PMU list~\cite{Intel2023sdm}, and iterate them to evaluate functions in PMU-Data. To overcome C3, we consider four methods to identify the special situation, and recover the secret data from different PMU counter values in \autoref{fig:overview}: (1) the point with the biggest value; (2) the point with the smallest value; (3) the point of demarcation where the value drops sharply; (4) the point of demarcation where the value rises steeply. We analyze instructions and design gadgets. For each gadget, we iterate all documented PMU events to implement PMU-Data, and utilize four methods to recover secret. We take the 8 bits as the size of secret data, which can be other size.
% We describe three main challenges to implement PMU-Data attack, and provide details to overcome C1 in \autoref{subsec:C1}. For C2 and C3, because the functions of PMU events are diverse, there are no common methods to identify vulnerable PMU events, and speculate the secret data from the special situations. To overcome C2, we refer to the documented PMU list~\cite{Intel2023sdm}, and iterate them to evaluate functions in PMU-Data attack. To overcome C3, we consider four methods to identify the special situation, and recover the secret data from different PMU counter values in \autoref{fig:overview}: (1) the point with the biggest value; (2) the point with the smallest value; (3) the point of demarcation where the data drops sharply; (4) the point of demarcation where the data rises steeply. All in all, we analyze instructions, and manually design gadgets. For each gadget, we iterate all documented PMU events to implement PMU-Data, and utilize four methods to recover secret. \textcolor{red}{In this paper, we take the 8 bits as the size of secret data, which can be other size.}

\subsection{Vulnerable Events and Trigger Instructions}
\label{subsec:results}

% Moreover, \texttt{L2\_RQSTS.DEMAND\_DATA\_RD\_HIT} event can achieve PMU-Data v2 for Meltdown and Zombieload on i7-6700, but fails on i7-7700. Different versions of micro-code may be the root cause.
Experiment results show that two kinds of instructions (\verb|DIV| and \verb|MOV|), and 40 PMU events are vulnerable to implement PMU-Data attack, in which the \texttt{ARITH.DIVIDER\_ACTIVE} event is for PMU-Data v1, and the other events are for PMU-Data v2. We test the transient execution vulnerabilities on the three processors, and evaluate case studies of PMU-Data. Intel i5-7300U is not vulnerable to Meltdown, so the PMU events for PMU-Data attack is just 36. Taking the i7-7700 as the example, we provide experiment results to implement Meltdown, Spectre and Zombieload in \autoref{table:results}, in which \verb|T| denotes the throughput (bytes per scond), and \verb|E| denotes the error rate (\%). We use two methods to handle the exception in Meltdown and Zombieload, and display their different results in throughput and error rate. For Zombieload, the secret data is protected by Intel SGX, and we find 13 PMU events available. Results in \autoref{table:results} demonstrate that Transactional Synchronization Extensions (TSX) can enhance the effectiveness of PMU-Data including throughput and accuracy for most PMU events.

% \subsection{Strength Analysis}
% \label{subsec:strength}
% The most related work is PMU-Spill~\cite{qiu2023spill}, which encodes the secret data into different execution paths of a gadget (can be regarded as a control flow-based attack), while PMU-Data encodes the secret data into the data traces (can be regarded as a data flow-based attack). There are two strengths for our PMU-Data attack. (1) PMU-Data can implement Spectre-PHT. PMU-Spill relies on a branch instruction to create different execution paths, which modifies the branch predictor unit (BPU)~\cite{evtyushkin2018branchscope} and makes it cannot implement Spectre-PHT. (2) There are more vulnerable gadgets to realize Spectre-PHT for PMU-Data. PMU-Data omits the \verb|JCC| instruction in PMU-Spill, directly uses just an instruction (\verb|DIV| and \verb|MOV|) to operate the secret data.
% %The \verb|DIV| and \verb|MOV| instructions are more common than \verb|JCC| instruction in normal programs.

% PMU-Data side-channel attack could be used to improve the effectiveness of transient execution attacks. Modern transient execution attacks utilize cache side-channel attacks to recover the secret data from the transient operations. Meanwhile, most researchers focus on memory requests and high-resolution timers to prevent from cache side-channel attacks~\cite{yu2019stt}. However, compared to side-channel attacks, our PMU-Data attack does not rely on the timing difference of any memory requests such as load and store operations. Therefore, our PMU-Data side-channel attack could not be prevented by any existing secure cache.

\subsection{Covert-Channel}

We assess the functionality of PMU-Data for constructing a covert-channel in the Intel i7-7700 machine. The two variants of PMU-Data attack can be employed to generate distinct values in the PMU counter, and recover the secret from the PMU values. We design a transmitting process. For example, if the PMU is configured with the \texttt{ARITH.DIVIDER\_ACTIVE} event, the transmitting process executes the division-by-zero operation to dispatch the signal \texttt{0}, and sends the signal \texttt{1} by modifying the divisor to a non-zero value. The receiving process and the transmitting process can either share the same logical core, or reside in SMT scenarios where the \texttt{ANY} Bit in \texttt{IA32\_PERFEVTSELx} should be set to allow the PMU recording events in all logical cores of the physical core. Our prototype verification speed was 12.44KB/s with the error rate 0\% in i7-7700. 
%\textcolor{red}{Results show that}
% The receiving process and the transmitting process can operate in two scenarios: one where they share the same logical processor core, and the other where they reside on two logical cores of the same physical core.

\begin{table*}[htb]
\centering
\linespread{1.05}
\caption{Experiment results for PMU-Data on the Intel i7-7700 machine}
\begin{threeparttable}
\label{table:results}
% \tiny
% \small
% \footnotesize
\scriptsize
  \begin{tabular}{c|c|c|c|cc|cc|cc|cc|cc}
 \hline
 \multirow{3}{*}{Event's Category} & \multirow{3}{*}{Number} & \multirow{3}{*}{PMU event} & \multirow{3}{*}{UMask} &  \multicolumn{4}{c|}{Meltdown}&\multicolumn{4}{c|}{Zombieload}&\multicolumn{2}{c}{Spectre-PHT}\\
  \cline{5-14}
 &&&&\multicolumn{2}{c|}{TSX} & \multicolumn{2}{c|}{No TSX}&\multicolumn{2}{c|}{TSX} & \multicolumn{2}{c|}{No TSX}&\multicolumn{2}{c}{No TSX}\\
 \cline{5-14}
&&&&T&E&T&E&T&E&T&E&T&E\\
\hline 
\multirow{1}{*}{ARITH}&0x14&DIVIDER\_ACTIVE&0x01&302&0&-&-&76169&0.33&12888&2.47&126.65&0\\
\hline 
\multirow{8}{*}{{L2\_RQSTS}}&\multirow{8}{*}{{0x24}}&{DEMAND\_DATA\_RD\_MISS}&0x21&1707&0&258&0&-&-&-&-&177.92&0\\ \cline{3-14}
&&{ALL\_DEMAND\_MISS}&0x27&1711&0&258&0.09&-&-&-&-&177.82&0\\ \cline{3-14}
&&{PF\_MISS}&0x38&1723&8.92&259&2.97&-&-&-&-&-&-\\ \cline{3-14}
&&{MISS}&0x3F&1721&1.98&258&0.99&-&-&-&-&178.82&0\\ \cline{3-14}
&&{DEMAND\_DATA\_RD\_HIT}&0xC1&-&-&-&-&-&-&-&-&177.62&0\\ \cline{3-14}
&&{PF\_HIT}&0xD8&1715&0&258&0&-&-&-&-&177.59&0\\ \cline{3-14}
&&{ALL\_DEMAND\_DATA\_RD}&0xE1&1720&2.97&257&3.96&-&-&-&-&177.77&0\\ \cline{3-14}
&&{ALL\_DEMAND\_REFERENCES}&0xE7&1726&1.98&-&-&-&-&-&-&178.62&0\\ \cline{3-14}
&&{ALL\_PF}&0xF8&1718&0&258&0.99&-&-&-&-&177.65&0\\ \cline{3-14}
&&{REFERENCES}&0xFF&1720&0.99&258&5.94&-&-&-&-&177.95&0\\
\hline 
{LONGEST\_}&\multirow{2}{*}{{0x2E}}&{MISS}&0x41&1720&0&258&0&1535&0&1457&0.05&177.81&0\\ \cline{3-14}
{LAT\_CACHE}&&{REFERENCE}&0x4F&1721&0&258&0.99&-&-&-&-&177.95&0\\
\hline
\multirow{1}{*}{{L1D\_PEND\_MISS}}&0x48&{PENDING}&0x01&1721&0&258&0.99&1535&0.14&1457&0.08&17.78&0\\
\hline 
\multirow{1}{*}{{L1D}}&0x51&{REPLACEMENT}&0x01&1719&8.91&257&5.94&-&-&-&-&177.86&0\\
\hline 
\multirow{1}{*}{{RS\_EVENTS}}&0x5e&{EMPTY\_CYCLES}&0x01&-&-&-&-&-&-&-&-&17.78&5.26\\
\hline
\multirow{1}{*}{{OFFCORE\_}}&\multirow{3}{*}{{0x60}}&DEMAND\_DATA\_RD&0x01&1719&0&258&0&1534&0.10&1458&0.10&17.81&0\\ \cline{3-14}
\multirow{1}{*}{{REQUESTS\_}}&&{ALL\_DATA\_RD}&0x08&1725&0&258&0&1533&0.02&1456&0.06&1778&0\\ \cline{3-14}
\multirow{1}{*}{{OUTSTANDING}}&&{L3\_MISS\_DEMAND\_DATA\_RD}&0x10&1726&0&259&0&1533&0&1458&0.02&17.78&0\\
\hline 
{UOPS\_}&\multirow{2}{*}{{0xA1}}&{PORT\_0}&0x01&-&-&-&-&-&-&-&-&17.74&0\\ \cline{3-14}
{DISPATCHED\_PORT}&&{PORT\_4}&0x10&-&-&-&-&-&-&-&-&177.90&0\\
\hline
&\multirow{6}{*}{{0xA3}}&{CYCLES\_L2\_MISS}&0x01&1720&0&258&0.99&1534&0.1&1458&0.08&-&-\\ \cline{3-14}
&&{CYCLES\_L3\_MISS}&0x02&1717&0&258&0&1532&0.14&1457&0.01&-&-\\ \cline{3-14}
{CYCLE\_}&&{STALLS\_L2\_MISS}&0x05&1719&0&258&0&1534&3.11&1458&0.02&-&-\\ \cline{3-14}
{ACTIVITY}&&{STALLS\_L3\_MISS}&0x06&1723&0&257&0&1533&1.78&1456&0.07&-&-\\ \cline{3-14}
&&{CYCLES\_L1D\_MISS}&0x08&1719&0&258&0&1534&0.15&1453&0.09&1.78&5.26\\ \cline{3-14}
&&{STALLS\_L1D\_MISS}&0x0C&1721&0&258&0&1531&0.41&1458&0.22&-&-\\
\hline
{EXE\_}&\multirow{2}{*}{{0xA6}}&{EXE\_BOUND\_0\_PORTS}&0x01&-&-&-&-&-&-&-&-&177.84&0\\ \cline{3-14}
{ACTIVITY}&&{1\_PORTS\_UTIL}&0x02&-&-&-&-&-&-&-&-&177.69&0\\
\hline
\multirow{2}{*}{{OFFCORE\_}}&\multirow{4}{*}{{0xB0}}&{DEMAND\_DATA\_RD}&0x01&1720&0&258&0&-&-&-&-&177.75&0\\ \cline{3-14}
&&{ALL\_DATA\_RD}&0x08&1717&0&258&0.99&-&-&-&-&177.09&0\\ \cline{3-14}
\multirow{2}{*}{{REQUESTS}}&&{L3\_MISS\_DEMAND\_DATA\_RD}&0x10&1723&0&259&0&1534&0.12&1457&0.33&177.67&0\\ \cline{3-14}
&&{ALL\_REQUESTS}&0x80&1724&0&258&0.99&-&-&-&-&17.89&0\\
\hline 
{UOPS\_EXECUTED}&\multirow{2}{*}{{0xB1}}&{THREAD}&0x01&-&-&-&-&-&-&-&-&177.86&0\\ \cline{3-14}
{DISPATCHED\_PORT}&&{STALL\_CYCLES}&0x02&-&-&-&-&-&-&-&-&177.75&0\\
\hline
\multirow{1}{*}{{L2\_TRANS}}&0xF0&{L2\_WB}&0x40&-&-&258&0.99&-&-&-&-&-&-\\
\hline
\multirow{1}{*}{{L2\_LINES\_IN}}&0xF1&{ALL}&0x1F&1719&0&256&0.99&-&-&-&-&177.01&0\\
\hline
\multirow{3}{*}{{L2\_LINES\_OUT}}&\multirow{3}{*}{{0xF2}}&{SILENT}&0x01&1722&0&258&1.98&-&-&-&-&1.78&0\\ \cline{3-14}
&&{NON\_SILENT}&0x02&1721&5.94&258&0&-&-&-&-&-&-\\ \cline{3-14}
&&{USELESS\_HWPF}&0x04&1719&0.99&-&-&-&-&-&-&-&-\\
\hline
% \multicolumn{4}{c|}{Average(PMU-Data v2)}&1720&1.09&258&1.03&1534&0.56&1457&0.09\\
% \hline
\end{tabular}
\begin{tablenotes}
\footnotesize
\item {M/S/Z: whether it can be utilized to implement Meltdown/Spectre/Zombieload. T: Throughput (bytes per second). E: Error rate (\%).}
\end{tablenotes}
\end{threeparttable}
\end{table*}

\section{MITIGATION AND DISCUSSIONS}
\label{sec:mitigation+discuss}
% \subsection{Mitigation}
\label{sec:mitigation}

We provide two alternative methods as the hardware-based countermeasures: 1) blocking the Performance Monitoring Unit (PMU); 2) adding rolling back mechanisms to the usage of PMU in transient environments. For the first method, it is effective but not practical. The second method is valid but can take a small performance overhead. From the software perspective, we recommend to inspect the program and prevent the vulnerable gadget for Spectre-PHT and PMU-Data, which can be implemented by software developers~\cite{Li2023analysis}.
\label{subsec:discuss}

The most related work is PMU-Spill~\cite{qiu2023spill}, which encodes the secret data into different execution paths of a gadget (can be regarded as a control flow-based attack), while PMU-Data encodes it into different data traces (can be regarded as a data flow-based attack). There are two strengths for our PMU-Data attack. (1) It can implement Spectre-PHT. PMU-Spill relies on a branch instruction to create different execution paths, which modifies the branch predictor unit~\cite{evtyushkin2018branchscope} and makes it cannot implement Spectre-PHT. (2) There are more vulnerable gadgets to realize Spectre-PHT. PMU-Data omits the \verb|JCC| instruction in PMU-Spill, directly uses just an instruction (\verb|DIV| and \verb|MOV|) to operate the secret data.
%The \verb|DIV| and \verb|MOV| instructions are more common than \verb|JCC| instruction in normal programs.

PMU-Data side-channel attack could be used to improve the effectiveness of transient execution attacks. Modern transient execution attacks utilize cache side-channel attacks to recover the secret data from the transient operations. Meanwhile, most researchers focus on memory requests and high-resolution timers to prevent from cache side-channel attacks~\cite{yu2019stt,zhang2024tscSurvey,yin2024detect}. However, compared to them, our PMU-Data attack does not rely on the timing difference of any memory requests such as load and store operations.
%Therefore, our PMU-Data side-channel attack could not be prevented by any existing secure cache.

Moreover, combined with two facts, the PMU-Data attack is powerful. (1) When implementing Meltdown and Zombieload, the PMU-Data attack can be realized using the signal mechanism (\verb|setjmp()| function) to handle exceptions. (2) Unprivileged attackers could access PMU values using transient execution vulnerabilities \cite{Weber2023Reviving}. Therefore, our PMU-Data attack does not rely on the support of TSX and root privilege, but threat many processors and scenarios.

\section{ACKNOWLEDGMENT}
This work was supported in part by the Beijing Natural Science Foundation (Grant No. 4242026), National Natural Science Foundation of China (Grant No. 62072263 and 62372258), the Fundamental Research Funds for the Central Universities (Grant No. 2023RC71), Tsinghua University Initiative Scientific Research Program. 

\section{CONCLUSION}
PMUs are designed to optimize applications' performance but can be utilized to leak secrets. In this study, we propose the PMU-Data attack to recover operands of instructions with PMU. Experiment results show that \verb|DIV| and \verb|MOV| instructions are vulnerable, and 40 PMU events are available to PMU-Data attack. We utilize the PMU-Data attack to implement Zombieload to steal secret data that is protected by Intel SGX on the Intel i7-7700 processor.

\bibliographystyle{IEEEtran}
\bibliography{refs}

\end{document}